\documentstyle[12pt]{article}

\makeatletter
\def\vereq#1#2{\lower3pt\vbox{\baselineskip1.5pt \lineskip1.5pt
\ialign{$\m@th#1\hfill##\hfil$\crcr#2\crcr\sim\crcr}}}
\makeatother

\begin{document}
\setlength{\unitlength}{1mm}
\textwidth 15.0 true cm
\textheight 22.0 true cm
\headheight 0 cm
\headsep 0 cm
\topmargin 0.4 true in
\oddsidemargin 0.25 true in

\def\beq{\begin{equation}}   \def\eeq{\end{equation}}

\newcommand{\gsim}{\lower.7ex\hbox{$\;\stackrel{\textstyle>}{\sim}\;$}
}
\newcommand{\lsim}{\lower.7ex\hbox{$\;\stackrel{\textstyle<}{\sim}\;$}
}

\newcommand{\ra}{\rightarrow}
\newcommand{\ve}[1]{\vec{\bf #1}}

\newcommand{\La}{\overline{\Lambda}}
\newcommand{\Lam}{\Lambda_{QCD}}
\newcommand{\re}[1]{Ref.~\cite{#1}}

\newcommand{\eq}[1]{Eq.\hspace*{.1em}(\ref{#1})}
\newcommand{\eqs}[1]{Eqs.\hspace*{.1em}(\ref{#1})}
\newcommand{\Eq}[1]{Eq.\hspace*{.1em}(\ref{#1})}
\newcommand{\Eqs}[1]{Eqs.\hspace*{.1em}(\ref{#1})}

\renewcommand{\Im}{\mbox {Im}\:}

\newcommand{\be}{\beta}
\newcommand{\ga}{\gamma}
\newcommand{\de}{\delta}
\newcommand{\al}{\alpha}
\newcommand{\as}{\alpha_s}

\newcommand{\GeV}{\,\mbox{GeV}}
\newcommand{\MeV}{\,\mbox{MeV}}

\def\lsim{\mathrel{\rlap{\lower3pt\hbox{\hskip0pt$\sim$}}
    \raise1pt\hbox{$<$}}}         
\def\gsim{\mathrel{\rlap{\lower4pt\hbox{\hskip1pt$\sim$}}
    \raise1pt\hbox{$>$}}}         

\begin{titlepage}
\begin{center}
\today     \hfill    SLAC-PUB-7769\\
~{} \hfill SU-ITP-98/13\\

\vskip .1in

{\large \bf The Hierarchy Problem and New Dimensions at a Millimeter}

\vskip .05in

Nima Arkani--Hamed$^*$, Savas Dimopoulos$^{**}$ and Gia Dvali$^{\dagger}$

$^*$ SLAC, Stanford University, Stanford, California 94309, USA

$^{**}$ Physics Department, Stanford University, Stanford, CA 94305, USA

$^{\dagger}$ ICTP, Trieste, 34100, Italy

\end{center}

\vskip .05in

\begin{abstract}
We propose a new framework for solving the hierarchy problem which does
not rely on either supersymmetry or technicolor.
In this framework, the gravitational and gauge interactions become 
united at the weak scale, which we take as the only fundamental
short distance scale in nature. The observed weakness of gravity 
on distances $\gsim$ 1 mm is due to the existence of 
$n \geq 2$ new compact spatial dimensions large compared to the weak scale.
The Planck scale
$M_{Pl} \sim G_N^{-1/2}$ is not a fundamental scale; its enormity 
is simply a consequence of the large size of the new dimensions.
While gravitons can freely propagate in the new dimensions, 
at sub-weak energies the Standard Model (SM) fields must be localized to a 
4-dimensional manifold of weak scale ``thickness" in the extra dimensions.
This picture leads to a number of striking signals for accelerator 
and laboratory experiments. For the case of $n=2$ new dimensions, planned
sub-millimeter measurements of gravity may observe the transition from
$1/r^2 \to 1/r^4$ Newtonian gravitation.
For any number of new dimensions, the 
LHC and NLC could observe strong quantum gravitational interactions.   
Furthermore, SM particles can be kicked off our
4 dimensional manifold into the 
new dimensions, carrying away energy, and leading
to an abrupt decrease in events with high transverse momentum $p_T \gsim$ TeV.
For certain compact manifolds, 
such particles will keep circling in
the extra dimensions, periodically returning, colliding with and
depositing energy to our four dimensional vacuum with frequencies
of $ \sim 10^{12}$ Hz or larger.
As a concrete illustration, we construct a model with SM 
fields localised on the 4-dimensional throat of
a vortex in 6 dimensions, with a Pati-Salam gauge symmetry $SU(4) \times SU(2)
\times SU(2)$ in the bulk. 

\end{abstract}

\end{titlepage}

\newpage

\section{Introduction}
There are at least two seemingly fundamental energy scales in nature,
the electroweak scale $m_{EW} \sim 10^3$ GeV and the Planck scale
$M_{Pl} = G_N^{-1/2} \sim 10^{18}$ GeV, where
gravity becomes
as strong as the gauge interactions.
Over the last two decades, explaining the smallness and radiative stability of
the
hierarchy $m_{EW}/M_{Pl} \sim 10^{-17}$ has been one of the
greatest driving forces
behind the construction of theories beyond the Standard Model (SM)
. While many different specific proposals for
weak and Planck scale physics have been made, there is
a commonly held picture of the basic structure of physics beyond the SM.
A new effective field
theory (e.g. a softly broken supersymmetric theory or technicolor)
is revealed at the weak scale, stabilizing and perhaps explaining
the origin of the hierarchy. On the other hand, the physics responsible for
making a
sensible quantum theory of gravity is revealed only at the Planck scale.
The desert between the weak and Planck scales could itself be populated with
towers of new effective field theories which can play a number of roles,
such as triggering dynamical symmetry breakings or explaining the
pattern of fermion masses and mixings.

In this picture, the experimental investigation of weak scale energies is
quite
exciting, as it is guaranteed to reveal the true mechanism
of electroweak symmetry breaking and stabilization of the hierarchy.
One can also hope that a detailed measurement of low energy parameters can
give valuable clues to the structure of effective field theories at higher
energies, perhaps even approaching the Planck scale. Nevertheless, it is
fair to say that in this paradigm, the thorough exploration of the weak scale
will never give a direct experimental handle on strong gravitational physics.

It is remarkable that such rich theoretical structures have been built on the
assumption
of the existence of two disparate fundamental energy scales, $m_{EW}$ and
$M_{Pl}$.
However, there is an important difference between these scales. While
electroweak
interactions have been probed at distances $\sim m_{EW}^{-1}$,
gravitational forces have not remotely been probed at distances $\sim
M_{Pl}^{-1}$:
gravity has only been accurately measured in
the $\sim 1 cm$ range. Our interpretation of $M_{Pl}$
as a fundamental energy scale (where gravitational interactions become strong)
is
then based on the assumption that gravity is unmodified over the 33 orders of
magnitude between where it is measured at $\sim$ 1 cm down to the Planck
length
$\sim 10^{-33}$ cm. Given the crucial way in which
the fundamental role attributed to $M_{Pl}$ affects our current thinking,
it is
worthwhile
questioning this extrapolation and seeking new alternatives to the
standard picture of physics beyond the SM.

In fact, given that the fundamental nature of the weak scale is an
experimental certainty, we wish to take the philosophy that $m_{EW}$ is the
only
fundamental short distance scale in nature, even setting the scale for the
strength of the
gravitational interaction. In this approach, the usual
problem with the radiative stability of the weak scale is trivially
resolved: the ultraviolet cutoff of the theory is $m_{EW}$.
How can the usual $(1/M_{Pl})$ strength of gravitation arise in such a
picture?
A very simple idea is to suppose that there are $n$ extra compact
spatial dimensions of radius $\sim R$. The Planck scale $M_{Pl (4+n)}$ of this
$(4+n)$ dimensional
theory is taken to be $\sim m_{EW}$ according
to our philosophy. Two test masses of mass $m_1,m_2$
placed within a distance $r \ll R$ will feel a gravitational potential
dictated by Gauss's law in $(4+n)$ dimensions
\begin{equation}
V(r) \sim \frac{m_1 m_2}{M_{Pl (4+n)}^{n+2}} \frac{1}{r^{n+1}}, \, (r \ll R).
\end{equation}
On the other hand, if the masses are placed at distances $r \gg R$, their
gravitational flux lines can
not continue to penetrate in the extra dimensions, and the usual $1/r$
potential is obtained,
\begin{equation}
V(r) \sim  \frac{m_1 m_2}{M_{Pl (4+n)}^{n+2} R^{n}} \frac{1}{r}, \, (r \gg R)
\end{equation}
so our effective 4 dimensional $M_{Pl}$ is
\begin{equation}
M_{Pl}^2 \sim M_{Pl (4+n)}^{2 + n} R^n. \label{fourplanck}
\end{equation}
Putting $M_{Pl (4+n)} \sim m_{EW}$ and demanding that $R$ be chosen to
reproduce
the observed $M_{Pl}$ yields
\begin{equation}
R \sim 10^{\frac{30}{n} - 17} \mbox{cm} \times \left(\frac{1
\mbox{TeV}}{m_{EW}}\right)^{1 + \frac{2}{n}}.  \label{radiuseq}
\end{equation}
For $n=1$, $R \sim 10^{13}$ cm implying deviations from Newtonian
gravity over solar system distances, so this case is empirically
excluded. For all $n \geq 2$, however, the modification of gravity
only becomes noticeable at distances smaller than those currently
probed by experiment. The case $n=2$ ($R \sim 100 \mu$m $-1$ mm) is
particularly exciting, since new experiments will be performed in
the very near future, looking for deviations from gravity in
precisely this range of distances \cite{aharon}.

While gravity has not been probed at distances smaller than a millimeter, the
SM gauge forces have certainly
been accurately measured at weak scale distances. Therefore, the SM particles
cannot freely propagate in the extra $n$ dimension, but must be localized to a
4 dimensional submanifold.
Since we assume that $m_{EW}$ is the only short-distance scale in the theory,
our 4-dimensional world
should have a ``thickness" $\sim m_{EW}^{-1}$ in the extra $n$ dimensions. The
only fields propagating in the
$(4+n)$ dimensional bulk are the $(4+n)$ dimensional graviton, with couplings
suppressed by the $(4+n)$ dimensional
Planck mass $\sim m_{EW}$.

As within any extension of the standard model at the weak scale,
some mechanism is needed in the theory above $m_{EW}$ to forbid
dangerous higher dimension operators (suppressed only by $m_{EW}$)
which lead to proton decay, neutral meson mixing etc. In our case,
the theory above $m_{EW}$ is unknown, being whatever gives a
sensible quantum theory of gravity in $(4+n)$ dimensions! We
therefore simply assume that these dangerous operators are not
induced. Any extension of the SM at the weak scale must also not
give dangerously large corrections to precision electroweak
observables. Again, there could be unknown contributions from the
physics above $m_{EW}$. However, at least the purely gravitational
corrections do not introduce any new electroweak breakings beyond
the $W,Z$ masses, and therefore should decouple as loop factor
$\times (m_{W,Z}/m_{EW})^2$, which is already quite small even for
$m_{EW} \sim 1$ TeV.

Summarizing the framework, we are imagining that the space-time is $R^4 \times
M_n$ for $n \geq 2$,
 where $M_n$ is an $n$ dimensional compact manifold of volume $R^n$, with $R$
given by eq. (\ref{radiuseq}). The $(4+n)$ dimensional Planck mass
is $\sim m_{EW}$, the only short-distance scale in the theory.
Therefore the gravitational force becomes comparable to the gauge
forces at the weak scale. The usual 4 dimensional $M_{Pl}$ is not a
fundamental scale at all, rather, the effective 4 dimensional
gravity is weakly coupled due to the large size $R$ of the extra
dimensions relative to the weak scale. While the graviton is free
to propagate in all $(4+n)$ dimensions, the SM fields must be
localized on a 4-dimensional submanifold of thickness $m_{EW}^{-1}$
in the extra $n$ dimensions.

Of course, the non-trivial task in any explicit realization of this
framework is localization of the SM fields. A number of ideas for
such localizations have been proposed in the literature, both in
the context of trapping zero modes on topological
defects\cite{defectUniverses} and within string theory.
In section 3, we will construct models of
the first type, in which there are two extra dimensions and, given
a dynamical assumption, the SM fields are localized within the
throat of a weak scale vortex in the 6 dimensional theory. We want
to stress, however, that this particular construction must be
viewed at best as an ``existence proof" and there certainly are
other possible ways for realizing our proposal, without affecting
its most important consequences.

 It is interesting that in our framework supersymmetry is no longer needed
from the low energy point of view for stabilizing the hierarchy,
however, it may still be crucial for the self-consistency of the
theory of quantum gravity above the $m_{EW}$ scale; indeed, 
the theory above $m_{EW}$ may be a superstring theory. 

Independently of any specific realization, there are a number of
dramatic experimental consequences of our framework. First, as
already mentioned, gravity becomes comparable in strength to the
gauge interactions at energies $m_{EW} \sim$ TeV. The LHC and NLC
would then not only probe the mechanism of electroweak symmetry
breaking, they would probe the true quantum theory of gravity!

Second, for the case of 2 extra dimensions, the gravitational force
law should change from $1/r^2$ to $1/r^4$ on distances $\sim 100
\mu$m-1 mm, and this deviation could be observed in the next few
years by the new experiments measuring gravity at sub-millimeter
distances\cite{aharon}. 

Third, since the SM fields are only
localized within $m_{EW}^{-1}$ in the extra $n$ dimensions, in
sufficiently hard collisions of energy $E_{esc} \gsim m_{EW}$, they
can acquire momentum in the extra dimensions and escape from our
4-d world, carrying away energy.\footnote{Usually in theories with
extra compact dimensions of size $R$, states with momentum in the
compact dimensions are interpreted from the 4-dimensional point of
view particles of mass $1/R$, but still localized in the 4-d world.
This is because the at the energies required to excite these
particles, there wavelength and the size of the compact dimension
are comparable. In our case the situation is completely different:
the particles which can acquire momentum in the extra dimensions
have TeV energies, and therefore have wavelengths much smaller than
the size of the extra dimensions. Thus, they simply escape into the
extra dimensions.} In fact, for energies above the threshold
$E_{esc}$, escape into the extra dimensions is enormously favored
by phase space. This implies a sharp upper limit to the transverse
momentum which can be seen in 4 dimensions at $p_T = E_{esc}$,
which may be seen at the LHC or NLC if the beam energies are high
enough to yield collisions with c.o.m. energies greater than
$E_{esc}$.  

Notice that while energy can be lost into the extra dimensions,
electric charge (or any other unbroken gauge charge) can not be lost.
This is because  the
massless photon is localized in our Universe and an isolated charge can
not
exist in the
region where electric field can not penetrate, so charges can not freely escape
into the bulk. In light of this fact, let us examine the fate of a 
charged particle kicked into the extra dimensions in more detail. 
On very general grounds (which we will discuss in more detail in section
3), the photon 
(or any other massless gauge field) can be
localized in our
Universe, provided it can only propagate in the bulk in the form of a massive
state with mass $\sim m_{EW}$, $m_{EW}^{-1}$ setting
the penetration depth of the electric flux lines into the extra
dimensions. In order for the localized photon to be  massless it is necessary
that the gauge symmetry
be unbroken at least within a distance  $\gg m_{EW}^{-1}$ from our
four-dimensional surface (otherwise the
photon will get mass through the ``charge screening", see section 3). 
As long as this condition is satisfied, the
four-dimensional observer will see an unbroken gauge symmetry with the right
4-d Coulomb law. Now, consider a particle with nonzero
charge (or any other unbroken gauge quantum number) 
kicked into the extra dimensions.
Due to the conservation of flux, an electric flux tube of the width
$m_{EW}^{-1}$ must be stretched between the escaping particle
and our Universe.  Such a string has a tension $\sim m_{EW}^2$  per unit
length. Depending on the
energy available in the collision, the charged particle will be 
either be pulled back to our
Universe, or the flux
tube will break into pieces with opposite charges at their ends. 
In either case, charge is conserved in the 4-dimensional world,
although energy may be lost in the form of neutral particles propagating 
in the bulk. Similar conclusions
can be  reached by considering a soft photon emission process\cite{oleg}.

Once the particles escape into the extra dimensions, 
they may or may not return to the 4-dimensional world, depending
on the shape and/or the topology of the $n$ dimensional compact manifold
$M_n$.  In the most interesting case, the particles orbit around the 
extra dimensions, 
periodically returning, colliding with and depositing 
energy to our 4 dimensional space with 
frequency $R^{-1} \sim 10^{27 - 30/n}$ Hz. This will lead to continuous 
``fireworks", which in the case of $n=2$ can give rise to $\sim$ mm displaced
vertices.

\section{Phenomenological and Astrophysical Constraints}
In our framework physics below a ~TeV is very simple: It consists
of the Standard Model together with a graviton propagating in 4+n
dimensions. Equivalently --in four dimensional language--our theory
consists of the Standard model
together with the graviton and all its Kaluza-Klein 
(KK) excitations recurring
once every $ 1/R $ , per extra dimension $n$. We shall refer to all
of them collectively as the ``gravitons'', independent of their
mass. Since each graviton couples with normal gravitational
strength $\sim{1/M_{Pl}}$ to matter, its effect on particle physics
and astrophysical processes is negligible. Nevertheless, since the
multiplicity of gravitons beneath any relevant energy scale $E$ is
$(ER)^n$ can be large, the {\it combined} effect of all the gravitons is
not always negligible and may lead to observable effects and
constraints. In this section we will very roughly estimate 
the most stringent of these constraints, mainly to show that our framework is
not grossly excluded by current lab and astrophysical bounds. Clearly,
a much more detailed study must be done to more precisely 
determine the constraints on $n$ and $m_{EW}$ in our framework.



Consider any physical process involving the emission of a graviton.
The amplitude of this process is proportional to $1/M_{Pl}$ and the
rate to $1/M_{Pl}^2$. Consequently, the total combined rate for
emitting any one of the available gravitons is
\begin{equation}
\sim {1 \over M_{Pl}^2} (\Delta E R)^n  \label{rate}
\end{equation}
where $\Delta E$ is the energy available to the graviton and the
last term counts the KK gravitons' multiplicity for n extra
dimensions. Using eq (\ref{fourplanck}) we can rewrite this as
\begin{equation}
 \sim {\Delta E^n \over m_{EW}^{2 + n}}\label{ratemodified}
\end{equation}
Note that the same result can be seen from the $4+n$ dimensional
point of view. The $m_{EW}$ suppressions of the couplings of the 
$4+n$ dimensional graviton are determined by 
expanding $g_{AB} = \eta_{AB} + h_{AB}/\sqrt{m_{EW}^{2 + n}}$, 
where $h_{AB}$ is the canonically normalised graviton in 
$4+n$ dimensions. Squaring this amplitude to obtain the rate yields precisely
the $m_{EW}$ dependence found above.
As a result, the  branching ratio for
emitting a graviton in any process is
\begin{equation}
\sim (\Delta E / m_{EW})^{2 + n}
\end{equation}
The experimentally most exciting (and most dangerous)
case has $m_{EW} \sim$ TeV and $n=2$. Of course, we must assume
that weak-scale suppressed operators giving proton decay,
large K-$\bar{K}$ mixing etc. are forbidden. Of the remaining lab constraints,
the ones involving the largest energy transfers $\Delta E$ (such as 
$\Upsilon$ and $Z$ decays) are most constrained. 
The braching ratio for graviton emisson in
Ypsilon decays is unobservable 
~ $\sim 10^{-8}$. For $Z$ $Z \rightarrow X +
graviton$ the branching ratio goes up to $\sim 10^{-5}$. Absence of
such decay modes puts the strongest laboratory constraints to the
scale $m_{EW}$ and/or n. Nevertheless, they are easy to satisfy, in
part because of their sensitivity to small changes in the value of
$m_{EW}$. Production of gravitons in very high energy collisions
will give the same characteristic signatures as the missing energy
searches, except for one difference: the missing energy is now
being carried by massless particles.

Next we consider astrophysical constraints. The gravitons are
similar to goldstone bosons, axions and neutrinos in at least one
respect. They can carry away bulk energy from a star and accelerate
its cooling dynamics. For this reason their properties are
constrained by the sun, red giants and SN 1987A. The simplest way
to estimate these constraints is to translate from the known limits on
goldstone particles. The dictionary that allows us to do that
follows from eq(\ref{ratemodified}) :
\begin{equation}
1/F^2 \leftarrow\rightarrow \Delta E^n/m_{EW}^{2 + n}
\end{equation}
relating the emission rate of goldstones and gravitons. Here $F$ is
the goldstone boson's decay constant. For the sun the available energy
$\Delta E$ is only a keV. Therefore, 
even for the maximally dangerous case $m_{EW}= 1$ TeV and $n=2$,
the effective $F$ is ~$10^{12}$ GeV, large enough to be totally safe
for the sun; the largest $F$ that is probed by the sun is $\sim 10^7$
GeV.

For red giants the available energy is $\sim $100 keV and the
effective $F
\sim 10^{10}$ GeV. This value is an order of magnitude higher than
the lower limit from red giants. Finally we consider the supernova
1987A. There, the maximum available energy per particle is presumed
to be between 20 and 70 MeV . Choosing the more favorable 20 MeV we
find an effective $F \sim 10^8$ GeV, which is smaller than the
lower limit of $10^{10}$ GeV claimed from SN 1987A. Therefore, the
astrophysical theory of SN 1987 A places an interesting constraint
on the fundamental scale $m_{EW}$ or/and the number of extra
dimensions n. The constraint is easily satisfied if $n>2$ or if
$m_{EW}> 10$ TeV. Of course, when the number of dimensions gets 
large enough so that 
$1/R \gsim 100$ MeV, (corresponding to $n \gsim 7$), none of the 
astrophysical bounds apply, since all the relevant temperatures
would be too low to produce even the lowest $KK$ excitation of the graviton.

Finally, although accelerators have not achieved collisions in the TeV energy
range where the most exotic aspects of the extra dimensions are revealed,
one may wonder whether very high energy cosmic rays of energies $\sim 10^{15}
-10^{19}$ eV (which in colliding with protons correspond to c.o.m. energies 
$\sim$ 1-100 TeV) have already probed such physics. However, the cosmic rays 
are smoothly accelerated to their high energies without any ``hard"
interactions, and they have dominantly soft QCD interactions with the protons
they collide with. Therefore, there are no significant constraints from very
high energy cosmic ray physics on our framework.

Having outlined our general ideas, some dramatic experimental consequences and
being reassured that existing data do not significantly constrain the 
framework, we turn to contructing an explicit model realising our picture,
with SM fields localised on the four-dimensional throat of a vortex in 6
dimensions.

\section{Construction of a Realistic Model.}

In this section we construct a realistic model incorporating the
 ideas of this paper. As stressed in the introduction, this should be
viewed as an example or an ``existence proof", since similar
constructions are possible in the context of field theory as well
as string theory. In particular one can change the structure and
dimensionality of the manifold, the localization mechanism, the
gauge group and the particle content of the theory without
affecting the key ideas of our paper. Furthermore, many of the
phenomenological consequences are robust and do not depend on such
details.

The space time is 6-dimensional with a signature $g_{AB} =$
(-1,1,1,1,1,1). The two extra dimensions are compactified on a
manifold with a radius $R \sim 1{\rm mm}$. We will discuss two
possible topologies: a two-sphere and a two-torus with the zero
inner radius. In both cases the key point is that the observable
particles ( quarks, leptons , Higgs and gauge bosons) are localized
inside a small region of weak-scale size equal to the inverse
cutoff length $\sim
\Lambda^{-1}$  and can
penetrate in the bulk only in form of the heavy modes of mass $\sim
\Lambda$. Thus for the energies $< \Lambda$  ordinary matter
gets confined to a four-dimensional hypersurface, our universe. The
transverse $x_5, x_6$ dimensions can be probed only through the
gravitational force, which is the only long-range interaction in
the bulk.

 There are several ways to localize the Standard Model  particles
 in our four-dimensional space-time. Here we consider the possibility that
localization is dynamical and the ordinary particles are ``zero
modes" trapped in the core of a four-dimensional vortex. This
topological defect, in its ground state, is independent of four
coordinates ($x_{\mu}$) and thus carves-out the four-dimensional
hypersurface which constitutes our universe.

 Consider first $x_5,x_6$ to be compactified on a two-sphere.
Define a six-dimensional scalar field $\Phi(x_A)$ transforming
under some $U(1)_V$ symmetry. We assume that $\Phi$ gets a nonzero
VEV $\sim \Lambda$ and breaks $U(1)_V$ spontaneously. The vortex
configuration is independent of
 the four
coordinates $x_{\mu}$ and can be set up through winding the phase
by $2\pi$ around the equator of the sphere:
\begin{equation}
  \Phi = \phi_{bulk} e^{i\theta}
\end{equation}
 where $2\pi > \theta > 0$ is  an azimuthal angle on the sphere and
$\phi_{bulk}$ is the constant expectation value that minimizes a
potential energy (modulo the small curvature corrections). Such a
configuration obviously implies two zeros of the absolute value
$\Phi$ on the both sides of the equator, which can be placed at the
north and the south poles respectively. These zeros represent the
vortex--anti-vortex pair of characteristic thickness $\sim
\Lambda^{-1}$. Since this size is much smaller than the separation
length $\sim {\rm 1 mm}$, vortex can be approximated by the
Nielsen-Olesen solution \cite{no}
\begin{equation}
\Phi = f(r)e^{i\theta}, ~~~f(0) =0, ~~~ f(r)|_{r >> \Lambda^{-1}}
\rightarrow
\phi_{bulk}
\end{equation}
where $0 < r < 2\pi R$ is a radial coordinate on the sphere, and an
anti-vortex corresponds to the change $\theta \to - \theta, r
\rightarrow 2\pi R -r$. If $U(1)_V$ is gauged the magnetic flux
will be entering the south pole and coming out from the north one.

 Since any closed loop on a two-sphere can be smoothly deformed to a
point, vortices on a sphere are not truly stable, and can
annihilate with anti-vortices if they encounter one another.
However the vortex anti-vortex pair are separated by a millimeter,
which is $10^{16}$ times their size; they therefore are, for all
practical purposes, stable. In addition there can be other
mechanisms of stabilization if other forces are involved (e.g.
repelling charges or currents flowing along the strings, etc.).

    Alternatively, compactification on a torus can lead to a truly
stable vortex. This is because a torus contains many
non-contractible loops, and the phase of $\Phi$ winding along such
a loop is topologically stable. Such a configuration is obtained
from the previously discussed two-sphere by identifying its poles
with a single point and subsequently removing this point from the
manifold. This manifold is then equivalent to a two-torus with zero
inner radius; it can carry topological charge and accommodate a
single vortex on it. The magnetic flux goes through the point that
was removed from the manifold. An observer looking at the south
pole will see the vortex with incoming flux. If he travels towards
the north pole along the meridian he will arrive to the same
vortex, since the poles have been identified, but will see it as an
anti-vortex since he will now be looking at the flux up-side down.

 Next, we come to the localization of  the standard
model particles on a vortex. We discuss the localization of
different spins separately.

\subsection{Localization of Fermions}

  Fermions can be trapped on the vortex as ``zero modes"  \cite{JR}
because of the index theorem \cite{indextheorem}. Consider a pair
of six-dimensional left-handed Weyl spinors $\Gamma_ 7 \psi,\xi =
\psi, \xi$. These, in terms of the four-dimensional chiral Weyl
spinors can be written as
\begin{equation}
\psi = (\psi_L, \psi_R),~\xi = (\xi_L, \xi_R)
\end{equation}
Assume now that this pair is getting a mass from coupling to the
vortex field:
\begin{equation}
 h \Phi \psi\xi + {\rm h.c.},
\end{equation}
where $h \sim \Lambda^{-1}$ has dimensions of inverse mass.
The six-dimensional Dirac equation in the vortex background is:
\begin{equation}
 \Gamma_A\partial^A\psi^{+} = h\phi_{bulk} e^{i\theta} \xi,
\end{equation}
and similarly for $\xi^+$. To look for solutions with localized
massless fermions we separate variables through the
angle-independent anzatz $\psi =
\psi(x_{\mu})\beta(r)$ and $\xi = \xi(x_{\mu})\beta(r)$, where $\mu
= 1,..4$, $\beta(r)$ is a radial scalar function in the 2 dimensional
compact space of $x_5$ and $x_6$. To be zero modes of the
four-dimensional Dirac operator, the spinors $\psi(x_{\mu})$ and
$\xi(x_{\mu})$ must satisfy
\begin{equation}
\Gamma_5e^{i\theta(-i\Gamma_5\Gamma_6)}\psi^+ \partial_r\beta(r) = h
\phi_{bulk}e^{i\theta}\xi,
\end{equation}
and similarly for $\xi^+$. Since $\psi^+$ and $\xi$ must be
eigenvalues of the $(-i\Gamma_5\Gamma_6)$ operator, they
automatically have definite four-dimensional chirality (say left
for the vortex and right for the anti-vortex). In this case the
normalizable wavefunction has the localized radial dependence
$\beta(r) = e^{- h\int^r_0f(r')dr'}$. Thus the vortex supports a
single four-dimensional massless chiral mode which can be
\begin{equation}
\psi_L + \xi_R^{+}
\end{equation}
 In general, as a consequence of the
 index theorem\cite{JR}\cite{indextheorem},  every pair of
 six-dimensional chiral fermions getting mass from the vortex
field, deposits a single zero mode of definite four-dimensional
chirality. Thus through the couplings to the vortex field one can
reproduce the whole set of the four-dimensional chiral fermions --
quarks and leptons-- localized on the submanifold.
 These localized  modes can get nonzero masses through the
usual Higgs mechanism. Let $\psi$ and $\psi'$ be the
six-dimensional chiral spinors (from different pairs) that deposit
two different zero modes on the vortex. These zero modes can get
masses through the couplings to a scalar field $H$
\begin{equation}
  H\psi\psi',
\end{equation}
 provided  it has a  nonzero expectation value in the core of the
vortex but vanishes in the bulk. The index is unaffected by the
existence of such a scalar since it has a zero VEV outside the
core.

\subsection{Localization of Higgs Scalars}

Now let us consider how the Higgs fields with non-zero VEVs can be
localized on the vortex. A massive scalar field can be easily
localized provided it has a suitable sign coupling to the vortex
field in the potential
\begin{equation}
  h'|\Phi|^2|H|^2
\end{equation}
If $h'>0$ , this contribution is positive in the bulk and zero in
the core. Thus $H$ will see the defect as an attractive potential,
which for a certain range of parameters can lead to a bound-state
solution. We will treat $H$ as the six-dimensional progenitor of
the Weinberg-Salam Higgs particle. Then the physically most
important case is when $H$ develops a non-zero expectation value in
the the defect. This is easily possible, provided an effective
mass$^2$ of the Higgs becomes negative in the throat of the vortex.
The simplest prototype potential of this sort is:
\begin{equation}
 (h'|\Phi|^2 - m^2) HH^+ +  c(HH^+)^2,~~~{\rm with}~~~ m^2,h', c >0,
\end{equation}
where $H$ is a six-dimensional scalar field. We will assume that
$h'\Phi_{bulk}^2 - m^2> 0$ and thus $H$ is zero in the bulk.
However, it can develop VEV in the vortex core. This does not
require any fine tuning and can be seen by examining the stability
of the trivial solution with respect to the small excitations
$H(x_5, x_6)e^{-i\omega t}$ in the vortex background. The analysis
is similar to the one of the superconducting cosmic string
\cite{Witten}. The linearized equation for small excitations is the
two dimensional Schrodinger equation
 \begin{equation}
-\partial_{5,6}^2 H + [h'f - m^2]H = \omega^2 H
\end{equation}
which certainly has a normalizable boundstate solution with
$\omega^2 < 0$ in a range of parameters. Thus $H $ becomes
tachionic in the core marking the instability of the trivial
solution $H = 0$. As a result $H$ develops an expectation value in
the throat of the vortex which decays exponentially for large $r$.
Since $H$ is a Higgs doublet there are the three massless Goldstone
modes localized on the vortex . These get eaten up by $W$ and $Z$
bosons through the usual Higgs effect. There is also a localized
massive mode, an ordinary Higgs scalar, which corresponds to the
small vibrations of the expectation value in the core $H(0)
\rightarrow H(0) +
h(x_{\mu})$. Such a vibration propagates in four-dimensions as the
ordinary massive scalar Higgs.

\subsection{Localization of Gauge Fields}

 There are several possible mechanisms  for localizing gauge
 fields on a vortex
 (or on any other topological defect)\footnote{ An alternate way
to localize massless gauge fields involves D-brane
constructions\cite{Dbranes}} through the coupling to the vortex
scalar. In general, a particle localized in such a way will not be
massless, unless there is a special reason such as the index
theorem for fermions and the Goldstone theorem or supersymmetry for
bosons. Here, we propose to localize massless gauge fields by
generalizing the four-dimensional confinement mechanism of
ref\cite{dvalishifman} (see also \cite{oleg}). Before discussing
how this mechanism is generalized to our case, it is instructive to
understand why the simplest approach fails. Consider the
$U(1)_{EM}$ electromagnetism in the presence of a thin vortex along
the z-axis. Let $\phi^-$ be a charged scalar field that develops an
expectation value and breaks $U(1)_{EM}$ spontaneously through the
Higgs mechanism. Now introducing a cross-coupling with the vortex
field
\begin{equation}
  (- a|\Phi|^2  + m^2) |\phi^-|^2 + b|\phi^-|^2 , ~~~ a,b,m^2 > 0
\end{equation}
and appropriately adjusting parameters (no fine tuning), we can
force $\phi^-$ to vanish in the vortex\footnote{An alternative
possibility is that $\phi^-$ is a vortex field itself, charged
under the electromagnetism, just as in the Abrikosov vortices in
 superconductors. In this case $\phi^-$ will vanish at the
origin for topological reasons.}; as a consequence, the tree-level
mass of the six-dimensional electric field coupled to it will also
vanish. Unfortunately,
 the four-dimensional photon trapped in this way does not remain
 massless. It
has a massive dispersion relation due to charge screening. This can
be understood as follows: Since the charged field $\phi^-$
condenses in the vacuum, the Universe is superconducting everywhere
except in the interior of the thin vortex.
 Two test charges placed at  different points $x_{\mu}$ and
$x'_{\mu}$ in the vortex will not interact by Coulomb's law; their
electric field polarizes the surrounding medium, and the field
lines end on the superconductor. 
As a result, the electric flux along the vortex dies-off exponentially
with (longtitudinal) distance within a characteristic length given
by the width of the vortex.

It is clear that for the localized gauge field to be massless  the
surrounding medium which repels the electric field lines should not
contain any charge condensate, otherwise all the field lines can be
absorbed by the medium. To construct such an example, consider a
thin planar-layer between two infinite superconductors. Two
magnetic monopoles located inside the layer
 interact through a long range magnetic
field. This is because the magnetic flux is repelled (or ``totally
reflected'') from the superconductor, since it contains no magnetic
charges on which the magnetic field lines can end. Consequently,
the magnetic flux is entirely contained inside the layer and, as a
result of flux conservation,
 the field lines spread according to  Coulomb's law.

In ref\cite{dvalishifman} a dual version of this mechanism --in
which the superconductor is replaced by a confining medium with
monopole condensation-- was suggested as a way to obtain massless
 gauge bosons localized on a sub-manifold.
Suppose that away from the vortex $U(1)_{EM}$ becomes a part of a
confining group which develops a mass gap $\sim \Lambda$. Then the
electric flux lines will be repelled by monopole condensation in
the dual Meisner effect; no images are created since there is no
charged condensate in the medium.

It is not difficult to construct an explicit four-dimensional
prototype model of this sort. It includes an $SU(2)$ Yang-Mills
theory with a scalar field $\chi$ in the adjoint representation,
plus a vortex field $\Phi$, which breaks some additional $U(1)_V$
symmetry and forms the string (for the present discussion it is
inessential whether $U(1)_V$ is global or gauged). The Lagrangian
has the form ($SU(2)$ indices are suppressed)
$$
{\cal L} = -\frac{1}{4g^2}{\rm Tr} G_{\mu\nu}G_{\mu\nu} +(D_\mu
\chi)^2
- \lambda(\chi^2)^2 -  \chi^2( h|\Phi|^2 - M^2) +$$
\beq
+ |\partial_\mu\Phi|^2
-\lambda'(|\Phi|^2 -\phi_{bulk}^2)^2\, ,
\eeq
where $G_{\mu\nu} $ is the ``gluon" field strength tensor, and
$h,\phi_{bulk}^2, M^2, \lambda, \lambda' $ are the positive
parameters and we assume $h\phi_{bulk}^2 > M^2$. In a certain range
of parameters the absolute minimum of the theory is achieved for
$\chi = 0$. In this vacuum, $\Phi$ develops the VEV $ \langle \Phi
\rangle
= \phi_{bulk}$ and
forms the vortex . Although $\chi$ is zero in the vacuum, it can
 acquire an expectation value inside the vortex, where
its mass$^2$ becomes negative, just like in the example with the
Higgs doublet considered above. In this case $SU(2)$ is broken to
$U(1)_{EM}$ on the string, but is restored outside. Inside the
string, two out of the three gluons acquire large masses of order
of $M$. The third gluon becomes a photon. Two degrees of freedom in
the $\chi$ field are eaten up by the Higgs mechanism, the remaining
degree of freedom is neutral. The massless degree of freedom in the
effective 1 + 1 dimensional theory on the string is a photon. It is
massless, since the $U(1)_{EM}$ gauge symmetry is unbroken
everywhere. On the other hand outside the vortex the photon becomes
a member of the nonabelian gauge theory, which confines and
develops a mass gap. Thus the photon can only escape from the
string in the form of a composite heavy ``glueball" with a mass of
the order of $\Lambda$ which we take to be the UV cutoff $\sim m_{EW}$. 
This guarantees that at low energies the
massless photon will be trapped on the string. The theory inside
the string is in the abelian 1 + 1 dimensional``Coulomb" phase.

 How is this mechanism generalized to our six-dimensional
case? Of course, we do not know how confinement
 works in a higher dimensional theory. Nevertheless, we believe and
 will  postulate that
 the  higher dimensional theory in the bulk will posses a
 mass gap $\sim \Lambda$ provided that:\\
 1) Outside the vortex the
standard model gauge group, in particular electromagnetism and
strong interactions, are extended into a larger nonabelian gauge
theory.\\
 2) There is no light
(with a mass below the cut-off scale $\Lambda$) matter in the bulk
enforced by general principles, such as Goldstone's theorem.\\
 3) The
tree-level gauge coupling blows up away from the vortex
\cite{oleg}.\\
 The latter condition can be satisfied e.g. if the
value of the gauge coupling is set by an expectation value of the
higgs field (or any function of it) which vanishes away from the
vortex. For instance, in the previous four-dimensional toy model
such a coupling is
\begin{equation}
\Lambda^{-2} {\rm Tr}  \chi^2 {\rm Tr}G_{\mu \nu}G^{\mu\nu}
\end{equation}

In summary, we presented one possible way for localizing particles
in our four-dimensional space-time. There are
 other possibilities within both field and string theory --such as
 D-brane constructions-- for accomplishing the same goal.

\subsection{A Realistic Theory}

 In this section we assemble the above ingredients to
  construct a prototype model  incorporating the ideas of this paper.
We embed the Standard Model in the Pati-Salam group $G =
SU(4)\otimes SU(2)_R\otimes SU(2)_L$ which is the unbroken gauge
group in the bulk. In addition, we introduce a $U(1)_V$ factor and
a singlet scalar field $\Phi$ charged under it. $\Phi$ develops an
expectation value and forms a vortex of thickness $\sim
\Lambda^{-1}$ in the compact 2-D submanifold spanned by $x_5, x_6$.
The interior of the vortex is our 4-dimensional space-time with all
the light matter confined to it. The only light particle
propagating in the bulk is the six-dimensional graviton.

The gauge group is spontaneously broken to $SU(3)\otimes U(1)_{EM}$
inside the vortex, by a set of six-dimensional scalar fields $\chi
= (15.1.1)$, $\chi'=(4.2.1)$ and $H=(1.2.2)$ which develop nonzero
VEVs only in the core of the vortex due to their interactions with
the $\Phi$ field. We assume a soft hierarchy $\chi' \sim \chi
\sim \Lambda \sim 10H \sim m_{EW}$. The crucial assumption is that in
the bulk the gauge group is strongly coupled and develops a mass
gap of the order of the cut-off. This, together with the fact that
$SU(3)\otimes U(1)_{EM}$ is unbroken everywhere guarantees that the
gluons and the photon are massless and trapped in our
four-dimensional manifold. $W^{\pm}$ and a $Z$ bosons are localized
as massive states.

 The matter fermions  are assumed to originate from the following
six-dimensional chiral spinors per generation:
\begin{equation}
  Q= (4,1,2),\bar Q= (\bar 4, 1,2), Q_c=(\bar 4, 2,1), \bar Q_c
=(4,2,1),
\end{equation}
which get their bulk masses through the coupling to the vortex
field
\begin{equation}
 h\Phi Q\bar Q  + h'\Phi^*Q_c\bar Q_c
\end{equation}
(where $h$ and $h'$ are parameters of the inverse cut-off size).
The index theorem ensures that each pair deposits a single chiral
zero mode which can be chosen as $Q_L + \bar Q_R^+$ and $ Q_{cL}^+
+
\bar Q_{cR}$ . These states get their masses through the couplings
to the Higgs doublet field which condenses in the core of the
vortex
\begin{equation}
  gH QQ_c  +  \bar g H\bar Q\bar Q_c \label{couplings}
\end{equation}

To avoid unacceptable flavor violations,the couplings in eqs(24)
should be flavor-universal. This can be guaranteed by some flavor
symmetry. Flavor violations must come from the ordinary Yukawa
couplings (see eq(\ref{couplings}) to be under control.

The theory presented here has rich and exotic phenomenology, thanks
to the existence of the extra dimensions. At energies above the
cutoff of a $\Lambda \sim$ TeV there is a plethora of particles
which can quite freely migrate and allow us to look into the extra
dimensions. Furthermore, naturalness requires that the migration
into the extra dimensions cannot be postponed much beyond the TeV
scale.

\section{Summary and Outlook}
The conventional paradigm for High Energy
Physics
--which dates back to at least 1974--  postulates that there are two
fundamental
scales , the weak interaction and the Planck scale. The large
disparity between these scales has been the major force driving
most attempts to go beyond the Standard Model, such as
supersymmetry and technicolor. In this paper we propose an
alternate framework in which gravity and the gauge forces are
united at the weak scale. As a consequence, gravity lives in more
than four dimensions at macroscopic distances --leading to
potentially measurable deviations from Newton's inverse square law
at sub-mm distances. The LHC and NLC are now even more interesting
machines. In addition to their traditional role of probing the
electroweak scale, they are quantum-gravity machines, which can also
look into extra dimensions of space via exotic phenomena such as
apparent violations of energy, sharp
high-$p_T$ cutoffs and the disappearance and reappearance of
particles from extra dimensions.


The framework that we are proposing changes the way we think about
some fundamental issues in particle physics and cosmology. The
first and most obvious change in particle physics occurs in our
view of
 the hierarchy problem. Postulating that the cutoff is at the weak
scale nullifies the usual argument about ultraviolet sensitivity,
since the weak scale now becomes the ultraviolet! The new hierarchy
that we now have to face, in the six dimensional case, is that
between the millimeter and the weak scales. This hierarchy is
stable in the sense that small changes of parameters have small
effects on the physics --so there is no fine tuning problem. There
is also no issue of radiatively destabilizing the mm scale by
physics at the weak cutoff. In this respect, our proposal shares
the same
 ``set it and forget it'' philosophy of the original proposal
supersymmetric standard model \cite{DG}. An important
and favorable difference is that the mm scale is not
 a Lagrangean parameter that needs to be stabilized by a symmetry,
such as supersymmetry. It is a parameter characterizing a solution,
the size of the two extra dimensions. It is not uncommon to have
solutions much larger than Lagrangean parameters; the world around
us abounds with solutions that are much larger than the electron 's
Compton-wavelength. A related secondary question is whether the
magnitude of the mm scale may be calculated in a theory whose
fundamental length is the weak scale. We have not addressed this
question which is imbedded in the higher dimensional theory. It is
amusing to note that if there are many new dimensions, their size
--given by eq (4)--
 approaches the weak scale and there is no large hierarchy.


Finally we come to the early universe. The most solid aspect of
early cosmology, namely primordial nucleosynthesis, remains intact
in our framework. The reason is simple: The energy per particle
during nucleosynthesis is at most a few MeV, too small to 
significantly excite gravitons. 
Furthermore, the horizon size is much larger than a mm
so that the expansion of the universe is given by the usual
4-dimensional Robertson-Walker equations. Issues concerning very
early cosmology, such as inflation and baryogenesis may change.
This, however, is not necessary since there may be just enough
space to accommodate weak-scale inflation and baryogenesis.

In summary, there are many new interesting issues that emerge in
our framework. Our old ideas about unification, inflation,
naturallness, the hierarchy problem and the need for supersymmetry
are abandoned, together with the successful supersymmetric
prediction of coupling constant unification
\cite{DG}. Instead, we gain a fresh framework which
allows us to look at old problems in new ways. Lagrangean
parameters become parameters of solutions and the phenomena that
await us at LHC, NLC and beyond are even more exciting and
unforseen.

\vspace{0.3cm}

{\bf Acknowledgments}: \hspace{0.2cm} We would like to thank 
I. Antoniadis, M.Dine, 
L. Dixon,  N. Kaloper, A. Kapitulnik, A. Linde, M. Peskin, 
S. Thomas and R. Wagoner for tuseful discussions. G. Dvali would like to thank
the Institute of Theoretical Physics of Stanford University for
their hospitality. NAH is supported by the Department of Energy under 
contract DE-AC03-76SF00515. SD is supported by NSF grant PHY-9219345-004.

\end{document}